      [1999/12/01 v1.4c Il Nuovo Cimento]
\documentclass{cimento}

\usepackage{graphicx} 

\title{FAVOR (FAst Variability Optical Registration) -- A Two-telescope
Complex for Detection and Investigation of Short Optical Transients}

\shorttitle{FAVOR (FAst Variability Optical Registration) --
Two-Telescope Complex}

\author{
G.~Beskin\from{sao},
V.~Bad'in,
A.~Biryukov, 
S.~Bondar, 
G.~Chuntonov, 
V.~Debur,
E.~Ivanov,
S.~Karpov,
E.~Katkova, 
V.~Plokhotnichenko, 
A.~Pozanenko, 
I.~Zolotukhin,
K.~Hurley,
E.~Palazzi, 
N.~Masetti, 
E.~Pian, 
L.~Nicastro,
C.~Bartolini, 
A.~Guarnieri,
A.~Piccioni, 
P.~Conconi,
E.~Molinari, 
F.~M.~Zerbi,
N.~Brosch,
D.~Eichler,
A.~Shearer, 
J.-L.~Atteia, 
\atque
M.~Boer
}
\shortauthor{G.~Beskin et al}

\instlist{
\inst{sao}
Special Astrophysical Observatory of RAS, Nizhnij Arkhyz, 
Karachai-Cherkessia, Russia, 369167
}
\PACSes{
\PACSit{95.55.Cs}{Ground-based ultraviolet, optical and infrared telescopes}
}
\begin{document}

\maketitle

\begin{abstract}
 An astronomical complex intended to detect optical
transients (OTs) in a wide field and follow them up with high time
resolution investigation  is described.
\end{abstract}


\begin{figure}
\includegraphics[angle=0,width=14cm]{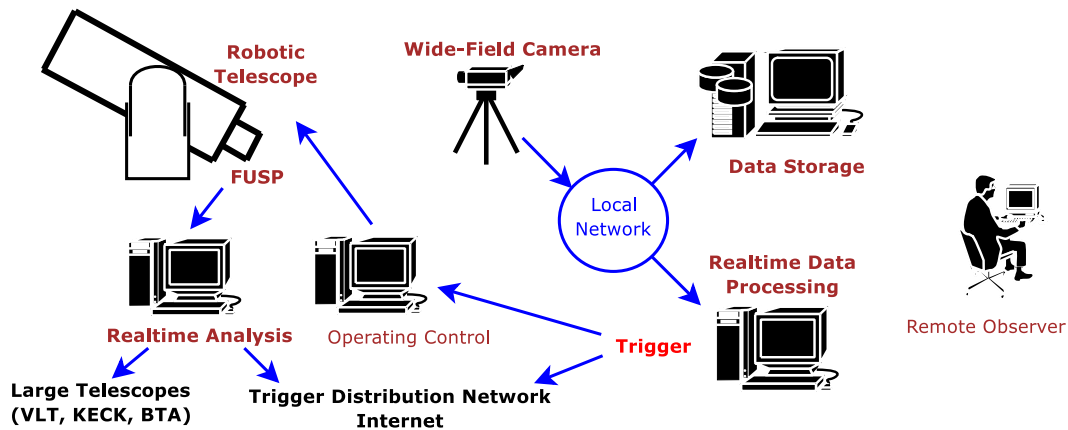}
\caption{General scheme of the FAVOR complex}
\label{complex}
\end{figure}

\begin{figure}
\includegraphics[angle=90,width=6.5cm]{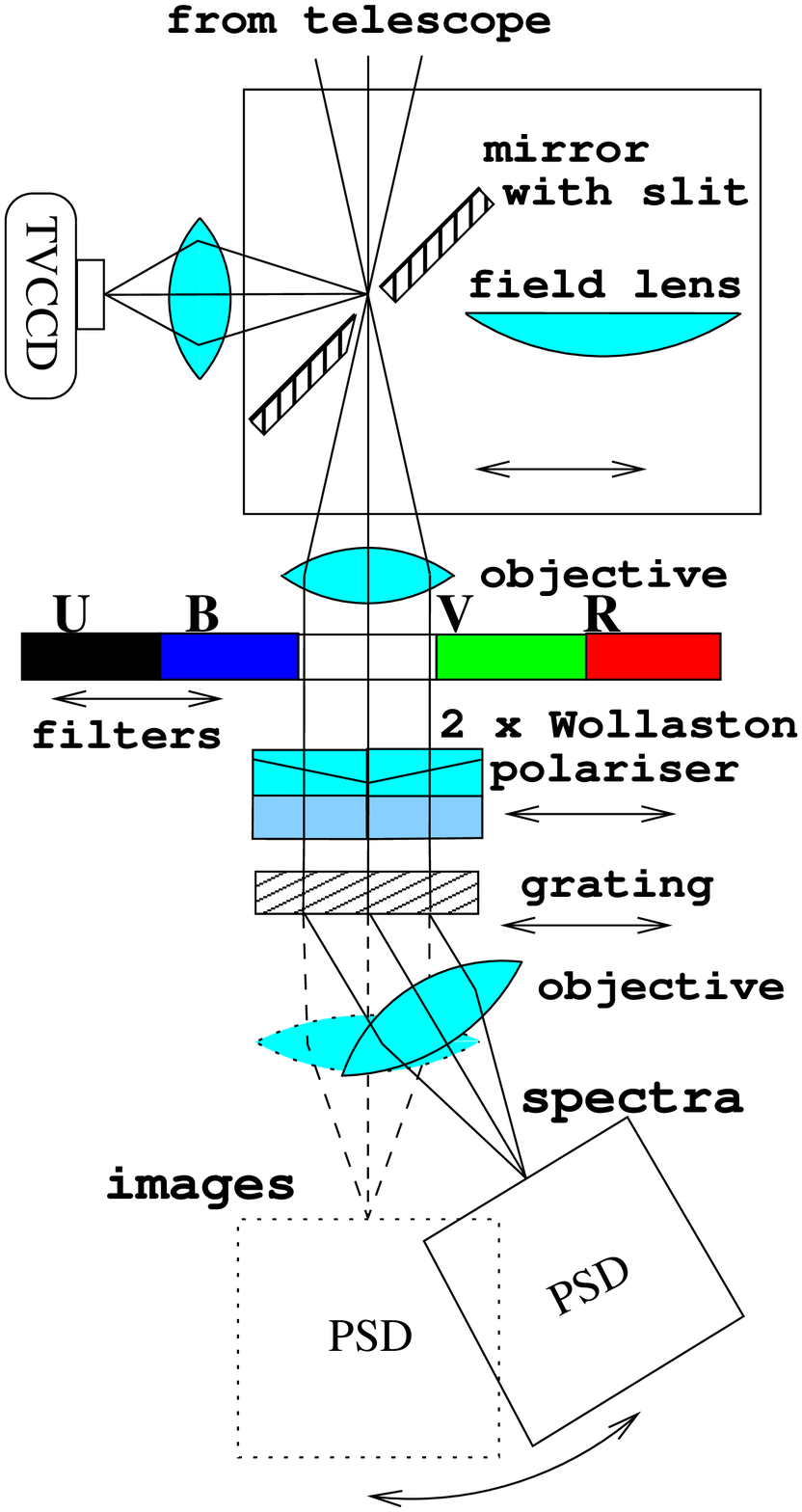}
\includegraphics[angle=0,width=6.5cm]{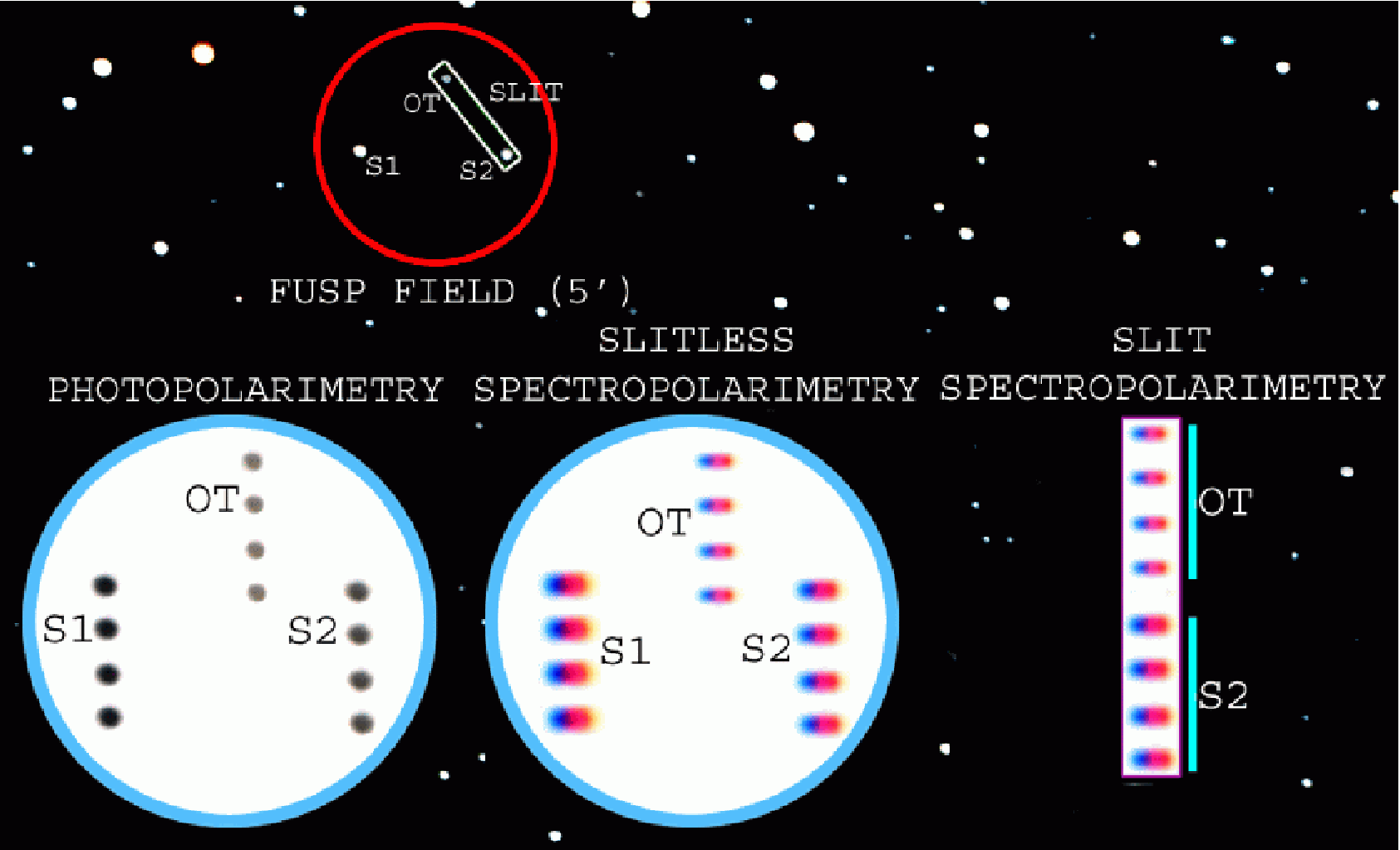}
\caption{FUSP principal scheme (left panel) and sketch of its field of view 
in various modes (right panel)}
\label{fusp}
\end{figure}


The nature of the Gamma-Ray Bursts (GRBs), the most powerful
events in the Universe, remain mysterious despite many
investigations in the last years.
It is clear that their progenitors are extremely compact stellar
objects, such as neutron stars (NS) or even black holes (BH),
either isolated or in binary systems. Their internal spatial
structure and dynamics, and the processes transforming their
energy into GRBs are inevitably reflected in the temporal
properties of bursts. While the full duration of the GRBs covers a
0.01 - 100 s range, the light curves of long ones show substructure in 80\% of
the cases, and for 66\% the time scales of the short ones variability are
shorter than 0.1 of their duration \cite{piran}. Moreover,
millisecond features are discovered in the light curves of some
long-duration bursts \cite{mcbreen}. At the same time, a number of
models predict the generation of considerable optical flux
simultaneous with the GRB, which can reach 8-12 magnitude
\cite{liang, eichler, wu}. This is especially important due to the
fact that for reasonable spectral assumptions the amount of
optical photons exceeds by factors of 100 - 1000 the number of
gamma-ray photons. The results of a unique quasi-synchronous
detection of a GRB optical transient (OT; the optical exposure was
started 22 s after the trigger)  confirm such a relation
\cite{akerlof}. Therefore, the search for, the detection and the
detailed high temporal resolution study of OTs accompanying GRBs
may bring decisive information on the nature of this phenomenon.

It is clear that observations such as described here must be
carried out independently of triggers created by space borne
telescopes with instruments equipped with wide fields of view,
comparable with those of satellite instruments ($>$ 100 deg$^2$).
This approach has been realized in the creation of wide-field
camera with high temporal resolution\cite{karpov}. Note, however,
that wide-field monitoring telescopes can not have objectives with
diameter large than 15-20 cm (due to the growth of optical
aberrations) and thus may be used only for the detection of OTs as
a source of triggers. These very wide field instruments must pass
the OT information on to the mid-sized telescope (50-200 cm
diameter) for detailed study. Two-telescope complexes must be
created\cite{potsdam}. In some sense, terrestrial automatic
optical telescopes receiving event triggers from space borne gamma
detectors form  two-telescope complexes similar to those described
above. However, due to time delays in the information transfer
through global networks it has never been possible to start
optical observations earlier than 20-60 s from the high-energy
trigger. Moreover, even UVOT, the dedicated optical telescope on
board the Swift satellite, starts observations only 30-100 s after
the gamma-ray photon detection.

Taking these into account, a set of requirements for a
two-telescope complex for the detection and investigation of fast
optical transients could be formulated:
formation of the trigger by the monitoring instrument in
real time within 0.5 - 1 s from detection or faster; 
minimal time delay for data transfer to the second-level
telescope (local network -- less than 0.5 s);
minimal pointing time towards the OT (slew speed faster
than 5-7 deg/s) -- 1-2 s;
maximal temporal resolution for both instruments --
better than 0.5 s for the monitoring one, better than 10$^{-3}$ s
for the second-level telescope;
extraction of maximum possible information - temporal,
spatial, spectral and  polarimetric - from each recorded source
photon by the second telescope instrument. Therefore, this is
conceived to be a panoramic spectropolarimeter;
fully automatic operation of the complex.

We proposed to assemble such a research complex and began its
development\cite{potsdam}.


The complex (see Fig. \ref{complex}) consists of a Wide-Field
Camera (WFC), a real-time data processing computer complex (RTDC)
and a robotic telescope (RT) equipped with a Fast Universal
Spectro-Polarimeter (FUSP).

The WFC\cite{zolotukhin} has an 15 cm objective with fast
(f$_\sharp$=1.2) focal ratio, which yields a wide field of view of
$16^{\circ}\times21^{\circ}$. It is equipped with a VS-CTT285-2001
TV-CCD camera (1280x1024 6.5 microns pixels) with a 7.5 Hz frame
rate (0.13 s exposure duration per frame). The data flow from the
camera is processed on several time scales simultaneously. The
shortest time scale is that of frame acquisition -- the transient
may be detected and classified if seen for 0.4 s - i.e., on three
successive frames (this aims to filter out the noise and moving
objects such as satellites and meteors).  The limiting magnitude
on this time scale is roughly 11.5 in a spectral band close to V.
On the second time scale, sums of 100 successive frames are
examined; this leads to a 14$^{\rm m}$ limit with 13 s temporal
resolution and allows the detection of slowly rising transients.

The WFC will operate in conjunction with a RT able to
automatically slew to the position of detected OT for detailed
studying. An example of such a telescope is the REM -- Rapid Eye
Mount \cite{zerbi}, which has a 60 cm mirror with a 4.8 m
effective focal length and is located at La Silla Observatory, in
Chile. Currently it can slew at a 5$^{\circ}s^{-1}$ rate and is
equipped with the REMIR near-infrared camera and ROSS (REM Optical
Slitless Spectrograph) instrument, therefore it can perform
detailed investigation of OTs on a time scale of a few seconds
after the trigger reception.

Such a RT equipped with the FUSP will allow detailed
investigations of OTs in spectropolarimetric mode and with high
time resolution; this would maximize the collected information
about each detected source photon.

The layout of FUSP is shown on the left panel of Fig.\ref{fusp}.
The device consists of an input  unit, which allows the choice of
three observing modes  (photopolarimetry, slitless and slit
spectropolarimetry), collimating  optics, a polarization unit (a
double Wollaston prism), a set of filters, a  dispersion element
(prism), a position-sensitive detector (PSD) with time  resolution
of 1 microsecond, and an acquisition system of type
"QUANTOCHRON-4-96"\cite{plokhotnichenko}.  In order to control the
observations, the FUSP is equipped with a TV-CCD guide.

 In PHOTOPOLARIMETRIC mode, the field lens forms  a 4-5 arcmin
 diameter image of the FOV in the collimator focal
plane. The PSD is placed on the same  axis with this lens, the
collimator, the polarizer and one of the filters in the set (U, B,
V, or R). The PSD records photons from each object in the FOV on
four simultaneous images, with different polarization plane
orientations. For an exposure time of one second  it is possible
to measure 30\% linear polarization (3 Stokes parameters
simultaneously) at a 3$\sigma$ level, for an object with
V$\approx$13.5 (S/N=10).

 In SLITLESS SPECTROPOLARIMETRIC mode, the PSD is placed on the same axis
with the prism and the collimator output lens. Images of four
spectra with different polarization plane orientations, for each
object in the FOV, are  projected onto the PSD photocathode. A
spectral range of 4000-7000\AA is covered with a spectral
resolution of 100-250\AA.
Under moderate seeing and sky brightness conditions it is possible
to measure a 10--15\% linear polarization at a 3$\sigma$ level for
an object with V$\approx$11.5-12 (S/N=10).

 In SLIT mode, a diagonal mirror with a slit size of 2$''$x4$'$
replaces the field lens, and the OT and its comparison star are
projected onto the slit. Photons of their four spectra are
recorded by the PSD. In this mode, detection limit is
V$\approx$13-13.5$^{\rm{m}}$ for the conditions described above.

 In order to change observing modes, a set of motors under computer control
 is used. An average duration for this mode change is a
few seconds.

The detector to be used in FUSP is the new generation of
Position-Sensitive Detector with high time resolution. This device
will be able to record individual photons with spatial resolution
of 10-20$\mu$m, measure their arrival times with accuracy of 10
nanoseconds, and accumulate these data continuously without any
gaps for several hours. Its quantum efficiency will be 20-35\% in
the 360-850 nm band while providing a large number of pixels
($10^{6}$) and event rates up to $10^5$ counts per second.

Simulated results of FUSP observations in different modes are
shown in the right panel of Fig.\ref{fusp}. FUSP is now in its
manufacturing stage, on the basis of the experience of development
of Multichannel Panoramic Photometer-Polarimeter which now is in
operation at the 6-meter telescope of SAO RAS \cite{debur,
plokhotnichenko}.

The selection of observation modes is based on the
brightness of the OT as measured by the WFC (on the last frame
used for its classification) and the prediction of its brightness
for the observations start time. The brightness of OTs may be
measured on a single frame (0.13 s exposure) or on a sum of 
100 frames (13 s); this yields detection limits in the
V band of 11.5 and 14 mag respectively.

The observation mode is selected during the pointing time (1-3 s):
for V $<$ 11.5 -- slitless
spectropolarimetry, and for V = 11.5-14 -- photopolarimetry, which
is followed by the slit spectropolarimetry after the determination
of the OT position.

To conclude, FAVOR will be able to detect and study with high
temporal resolution OTs as faint as 14$^{\rm m}$ in 1-60 s after
their detection.

\acknowledgments

This work was supported by grants of CRDF (No. RP1-2394-MO-02),
RFBR (No. 04-02-17555), and INTAS (04-78-7366). S.K. thanks
Russian Science Support Foundation for support. G.B. thanks the
Cariplo Foundation for a scholarship, and  Merate Observatory for
hospitality.

\end{document}